# Designing a Collaborative Research Environment for Students and their Supervisors (CRESS)

Vita Hinze-Hoare
School of Electronics and Computer Science
University of Southampton
United Kingdom
2007

**Abstract**
*In a previous paper the CSCR domain was defined. Here this is taken to the next stage where the design of a particular Collaborative Research Environment to support Students and Supervisors ( CRESS) is considered. Following the CSCR structure this paper deals with an analysis of 13 collaborative working environments to determine a preliminary design for CRESS in order to discover the most appropriate set of tools for its implementation.*

## Introduction
In the previous paper a definition of the CSCR environment was provided which demonstrated that the CSCW and CSCL environment by themselves are not rich enough to encompass the requirements of collaborative research. An additional five research spaces were identified as necessary components for a CSCR environment. In this paper the application of the CSCR domain to the specific needs of supporting collaborative research students and their supervisors (CRESS) will be analysed with a view to obtaining the specific set of tools required within those categories for the design of the CRESS interface.

## Analysis of appropriate categories and tools for CRESS
Following the methodology of Lindgaard *et al* (2006), 13 working environments and three learning environments have been analysed. In order to determine the most relevant tools which might be applied in the construction of a CSCR environment the analysis has been based upon an assessment of advantages and disadvantages for each tool set with reference to the needs of collaborative research. The final set of tools is instanced in table 16 which summarises the toolset to be employed initially in the new CRESS interface.

## E-laboratory Analysis
Eleven different CSCL e-laboratory interfaces were analysed with a view to determining the range of tools available and a classification groups into which those tools belonged.
**Argles *et al* (2006)** have an e learning laboratory called "*CECIL*" which is designed to enable pairs of students to collaborate in the writing of program code. The interface allows them to see the output of their work as well as a simulated LED display.
**Bachler *et al* (2004)** employ an instant message client called "*Buddy Space*" to facilitate multiple views of collaborative workgroups together with information about the location, attendance and recording of virtual meetings.
**Baker *et al* (2002)** have analysed commercial real time distributed groupware called "*Groove*". This contains a real time collaborative workspace based upon text and voice chat.
**Berger *et al* (2001)** have set up a CSCL environment called " *Le Scenario*" to support community health projects. Their environment stimulates social interaction in a face to face web based learning space which provides access to a range of knowledge sources.
**Dalziel (2003)** has developed an e learning environment called "*Learning Design*" together with a learning activity management system *LAMS* which facilitates student run time activity and teacher run time monitoring.
**Harper *et al* (2004)** have created a three dimensional virtual learning environment referred to as the experimental team room "*ETR*". This allows participants to move freely around a virtual room set up like a standard meeting room. It also includes an electronic meeting assistant (*EMMA*) which provides a human face to interact with and to accomplish basic task in the environment.

**Hosoya *et al* (1997)** of Japan have developed 3D virtual reality environment called "*HyClass*" based on "*CORBA*" which allows the user to walk around, pick up objects, move them from place to place and share them with other users all in the form of representatives or avatars within the environment.
**Kligyte *et al* (2001)** have designed an interface named "*Fle3*" for the *ITCOLE* project which looks and acts much like a standard VLE but that allows a limited degree of shared working.
**Miao *et al* (2005)** have been employing a CSCL tree-based script authoring tool called "I*MS-LD*" which can be used collaboratively to create learning scenarios for students.
**Pekkola (2003)** uses the "*VIVA*" interface to support peripheral awareness in a 3D virtual environment. This allows the use of common artefacts for framing activities in workplaces.
**Walters' *et al* (2006)** "*MGrid*" framework provides a method for learning distributed computing. Although not properly a collaborative environment it does enable the rapid prototyping of distributed systems within a basic browser framework to enable security through a sandbox approach. This is designed for many machines to do the work of one.
**Liccardi *et al* (2006)** has produced a wiki system to improve workspace awareness to advance effectiveness of co-authoring activities. This co-authoring wiki system (CAWS) is designed to improve the user's response to document development and to extend the area of workspace awareness.
**Sim *et al* (2005)** have discussed a Web/Grid Services approach for a Virtual Research Environment (VRE). They are working on CORE which is a project to develop a VRE to enable orthopaedic surgeons to collaborate in the design, analysis and dissemination of experiments. Individual user spaces are supplemented by templates for standard documents, a database for experiments, access to e-print archives and a limited discussion facility between collaborators.

## VLE Analysis

A number of these tools are built in to standard VLE interfaces and may well be useful in the CSCR environment. Three VLEs have been considered: Blackboard/WebCT, Moodle and Elgg. These have been incorporated into table 1. The VLE's section display a range of social interaction tools which are not particularly evident in the other sections of that table. They also contain community creation and access authorisation tools which are useful to set the boundaries of the collaborative group and provide a secure environment for the exchange of ideas.

Web 2.0 tags which are a community device to allow the marking of content for the purpose of facilitating rapid search may only have a limited use in this environment as the utility of tags is proportional to the number of users within the community group. In large communities such as flickr.com tags are immensely useful whereas in the much smaller groups of the CSCR environment their usefulness would be diminished.

Friend file sharing and blogging are both methods for making data available to a wider audience and would both be considered useful tools in a CSCR environment. Blogging can also play the dual role of a journal or log which can either be public or private, facilitating the process of reflection within the community. RSS feeds provide a central point for the aggregation of widely published data sources and provide a mini customisable portal which can focus the interests of a particular research group.

Peer review assistance would be useful in a number of areas. The provision of a database of academic peers and papers would assist research, but this may be difficult to provide internally to a CSCR environment. A fuller database is usually available on dedicated websites such as ACM, BCS, arxiv etc. which perform this kind of role more adequately. All that may be required in the CSCR environment is a link to the external databases. Finally public spaces and private spaces can both be useful in this environment where the former allows individual contributors to formulate their work prior to sharing, and public spaces allow the canvassing of opinion of a wider audience to raise public issues and survey opinion.

Table 1 shows the various toolkit elements employed by each of the interfaces and VLEs mentioned above, where the X mark in the table indicates that the feature is implemented in the e-laboratory. The results show that apart from Login and Access tools the most utilised tools are text chat and file depository (8 out of 16). The second most popular tools are scheduling and forum with 7 out of 16, and the third most popular tools are the help pane, the message board and the collaborative working window (6 out of 16).

| | CATEGORY SPACES | TOOLS | Argles SIM20 | Bachler "Buddy Space" | Baker et al "Grove Space" | Berger | Dalziel "LAMS" | Harper et al "ETR" | Hosoya et al "HyClass" | Kligyte et al "Fle3" | Miao et al "IMS-LD" | Pekkola "VIVA" | Walters et al "M-Grid" | Liccardi CAWS: A co-authoring Wiki | Sim et al Web/Grid Services VRE | Blackboard and WebCT | Moodle | Elgg | total |
|---|---|---|---|---|---|---|---|---|---|---|---|---|---|---|---|---|---|---|---|
| CSCR / CSCL / CSCW | Administration Space (including security tools) | Login | x | x | x | x | x | x | x | x | x | x | x | x | x | x | x | x | 16 |
| | | Access/authorisation Tools | x | x | x | x | x | x | x | x | x | x | x | x | x | x | x | x | 16 |
| | | Recording /Replay Facility | | x | | | | | | | | | | | | | | | 1 |
| | | Instant Messaging Recording | | x | | | | | | | | | | | | | x | | 2 |
| | | Assistive Agent | | | | | | x | | | | | | | | | | | 1 |
| | | Help Pane | x | | x | | | | x | | | | | | | x | x | x | 6 |
| | | Information Link Map | | x | | | | | x | x | | | | | | x | | | 4 |
| | | Scenario/Control flow Tools | | | | | | | | | x | | | | | | | | 1 |
| | Communication space (including Identification space) | Text/ Chat | x | | x | | x | x | x | | x | | | | | x | | x | 8 |
| | | Audio/Voice | | | x | | x | x | | x | | | | | | | | | 4 |
| | | Still Picture | | x | | | | | x | | | | | | | x | x | x | 5 |
| | | Video | | x | | | | x | x | | | | | | | | | | 3 |
| | | Instant Messaging | | x | | | | | | | x | | | | | | x | | 3 |
| | | Forum | | | x | | | | x | | | | x | x | x | x | x | x | 7 |
| | | Message Board/News | | | | x | | | | x | x | | | x | | x | x | x | 6 |
| | | Avatar (Representations) | | | | | | | x | | | | | | | | | | 1 |
| | | Presence Indicator/Information | | x | x | | | | | | x | x | | | | | x | | 5 |
| | | Location Identifier | | x | | | x | | | x | | | | | | | | | 3 |
| | | Focus Indication | | x | | | x | | x | | | | | | | | | | 3 |
| | | Participant Data | | x | | | | | | | | | | x | x | | x | x | 5 |
| | Scheduling space | Scheduling Tool | | x | | x | | x | | x | | x | | | | x | x | | 7 |
| | | Task Setting | | x | | | | | | | x | | | | | x | x | | 4 |
| | | Task Monitoring | | x | | x | | | | x | | | x | | | x | | | 5 |
| | Shared working space | Whiteboard | | | x | | | | x | x | x | | | | | x | | | 5 |
| | | Collaborative Working Window | x | | x | | | x | x | | | | x | | | | | x | 6 |
| | | 3D Environment | | | | | | x | x | | | | | | | | | | 2 |
| | Product Space | Output Window | x | | | | | | | | | | x | | | | | | 2 |
| | | Simulations | x | | | | | | | | | | | x | x | | | | 3 |
| | Reflection Space | Reflective Journal/Private | | | | | x | | | | | | x | | | | x | x | 4 |
| | Social Interaction Space | Community Creation | | | | | | | | | | | x | | | | | x | 2 |
| | | Tags (marking Content) | | | | | | | | | | | x | | | | | x | 2 |
| | | Friend (file sharing) | | | | | | | | | | | x | | | | | x | 2 |
| | | Blog (Public + Private) | | | | | | | | | | | x | x | | | x | x | 4 |
| | | RSS feed to centralize data | | | | | | | | | | | | | | x | x | | 2 |
| | Assessment / Feedback Space | Assessment | | | | | | | | | | | | | | x | x | | 2 |
| | | Feedback | | | | | | | | | | | | | | x | x | x | 3 |
| | Supervisor Space Space | Private area for tutors | | | | | | | | | | | | | | x | x | x | 3 |
| | Knowledge Space | Contribution Database | | | | x | | | x | x | | | | x | x | | | | 5 |
| | | Academic database | | | | | | | | | | | | | x | | | | 1 |
| | | Depository | | | | x | | | x | x | x | | | | x | x | x | | 8 |
| | | PowerPoint Slides | | x | | | x | | | | | | | | | | x | x | 4 |
| | Privacy Space | Private Space | | | | | | | | | | | | | x | x | x | x | 4 |
| | Public Space | Public information space | | | | | | | | | | | | | | x | x | x | 3 |
| | Negotiation Space | Peer Review assistance | | | | | | | | | | | | x | x | | x | | 3 |
| | Publication Space | Schemas/Templates | | | | | | | | | | | | x | | | | x | 2 |
| | | Publishing assistance | | | | | | | | | | | | x | x | x | x | | 4 |

**Table 1: Analysis of tools available to diverse e-learning systems**

## Categorisation and Selection of the appropriate tools for CRESS

Following Hinze-Hoare (2006) The CSCR domain was analysed and the services provided were factored into a number of distinct logical categories as follows:

- Administration
- Communication
- Scheduling
- Sharing
- Product

**CSCW**

- Reflection
- Social
- Assessment/Feedback
- Supervisor

**CSCL**

- Knowledge
- Privacy
- Public
- Negotiation
- Publication

**CSCR**

Forty-six tools in these 14 categories have been identified as being utilised within CSCW, CSCL and CSCR environments. The tools are now examined for inclusion in the CRESS interface.

The process that it used will be a critical analysis involving a determination of the advantages and disadvantages of the utility for each tool. This will be done on a category by category basis until an appropriate set of tools is arrived at for CRESS. Each of the tools required within these primary categories, see Table 1 will be considered in detail now.

Those tools which are specific to CSCW will be considered in tables 2 to 6.

## Administration Space Tools (Security, Recording and logging tools)

| Description of Tools | Advantages | Disadvantages |
|---|---|---|
| Access/Authorisation Tools | o Limits availability to authorised user<br>o Limits access to specific areas<br>o Accountability of actions through tracking | o Usual security overheads |
| Login | o Tracking | o |
| Recording/Replay of text (instant messaging) | o Allows detailed analysis of ideas and content<br>o Can be attended to at own time | o Low storage overheads |
| Recording/Replay of audio | o Medium level of communication | o Medium storage overheads |
| Recording/Replay of video | o Non verbal communication<br>o Highest level of communication | o High storage overheads |
| Recording/Replay of computer actions | o Enhances visualisation and demonstration roles | o Low storage overheads |
| Help Pane | o A simple statement of important facts of the operation of the interface easily accessible | o Help panes can obscure parts of the interface |
| Information link map | o A menu of help in form of graphical display of links | o |
| Assistive agent | o Interacts with the user to provide artificial intelligent help | o Requires a high degree of sophisticated programming expertise<br>o Will take a long time to develop |
| Scenario control flow tools | o | o Specific to a particular task |

**Table 2 Critical evaluation of Administration tools**

These encompass Help, Security and Recording tools. Help tools include the use of the interface control, information link maps and a simple help pane. Additionally an assistive agent could be used employing a degree of artificial intelligence to provide a higher degree of help (setting up the scenario etc.) Recording tools include mechanisms for recording communication transactions both for the purposes of reviewing information and for logging and validation. This includes security for **authentication authorisation and accounting** in the CSCR environment. This would require the implementation of basic methods such as login and password procedures etc.

Audio recording/replay can be included as a subset of video recording/replay which includes both moving pictures and sound. The replay of computer actions (recordable macros) which store keyboard presses and mouse movements can be useful for demonstration purposes particularly in a whiteboard environment.
Both the help pane and the information link map would provide useful help features in an accessible format with the link map providing a graphical index for quick access. These can be base on a simple hypertext (HTML) system and should be easy to set up and administer within the CSCR environment.

The assistive agent is deemed to require too high a programming outlay to merit the advantages to be obtained. The scenario control flow tools are specific to particular needs and don't apply to a generic CSCR interface.

## Communication Space Tools

| Description of Tool | Advantages | Disadvantages |
|---|---|---|
| Text/chat | o Can be recorded easily<br>o More concise<br>o Small file sizes | o Higher degree of effort<br>o Typing skills required<br>o Absence of verbal communication<br>o Absence of non verbal communication<br>o Requires appointed time |
| Audio | o Can be recorded easily<br>o Immediacy<br>o Easy of use | o Absence of non-verbal communication<br>o Increased file sizes<br>o Requires appointed time |
| Still Picture | o Quick visual identification | o Slight increase in memory requirements |
| Video | o Easily recordable<br>o Immediacy<br>o Ease of use | o Highest file sizes<br>o Requires appointed time |
| Instant Messaging | o Instant alert to online user<br>o Recordable<br>o Can be used synchronously and asynchronously | o Can distract from other work<br>o Higher degree of effort<br>o Typing skills required<br>o Absence of verbal communication<br>o Absence of non verbal communication |
| Forum | o Asynchronous communication<br>o Recordable<br>o Track individual ideas through a thread | o Lacks immediacy<br>o Higher degree of effort<br>o Typing skills required<br>o Absence of verbal communication<br>o Absence of non verbal communication |
| Message board/News | o One to many communication<br>o Useful news distribution | o One way communication<br>o Lacks immediacy<br>o Higher degree of effort<br>o Typing skills required<br>o Absence of verbal communication<br>o Absence of non verbal communication |
| Avatar (Representation of Participants) | o Quick visual identification<br>o Expression of non-verbal communication | o Large file size<br>o Higher overheads in operating cost |
| Presence Indicator | o Knowledge of Participants' presence<br>o Low overheads | o |
| Location Identifier | o Provides spatial indication | o Not required in a non geographical environment |
| Focus Indicator | o Identifies the speaker in synchronous communication | o |
| Participant Data | o Indicates name and other information of each participant | o May include irrelevant data |
| Email | o E-record | o Spam<br>o Higher degree of effort<br>o Typing skills required |

|  |  | o Absence of verbal communication<br>o Absence of non verbal communication<br>o Not dedicated to CSCR |
|---|---|---|
| Voice message | o Recordable | o One way communication<br>o Lacks immediacy<br>o Absence of non verbal communication |

**Table 3: Critical evaluation of communication tools**

These are for the exchange of information between collaborators. They include both synchronous and asynchronous tools.

A range of synchronous and asynchronous tools will be required. Synchronous tools are text chat, audio/voice, video and instant messaging. There is a clear preference for video over audio and audio over text as this mirrors more closely normal communication. However implementation is shown to be the reverse with the majority of interfaces using chat and the smallest number using video. It is likely that this is due to a higher degree of complexity in implementing the video feature which has resulted in this trend.

The implications for this are that for the CRESS environment the tools which bring the greatest degree of communication (**video**) are preferred over the more onerous to use tools (**chat**). In order to be rigorous it may be necessary to set up a communication interface with all three methods and determine which is the most widely used and in which circumstances. Asynchronous tools include **forum** which allows threads of conversation to be maintained and a simple message board that allows one- off news items to be posted.

The three tools for synchronous communication: Chat, Audio, Video have a distinct ordering in terms of ease of use. Audio and Video are easier to use than chat (which requires typing). Furthermore they have a distinct ordering in terms of the amount of information that can be communicated. Audio can communicate more information than chat for the same amount of human effort, and correspondingly Video can communicate more information than audio for the same amount of human effort. For this reason Video will be preferred over Audio and Audio over chat. The CRESS environment should contain Video communication, which can fall back to audio only if required. It is debatable whether a chat facility is needed in these circumstances. It is acknowledged that some forms of chat e.g. MSN are more popular than some forms of audio e.g. Skype. However, there are a number of reasons for this including the longer establishment of MSN, the 'zero cost for all users' universality of MSN whereas Skype is not free for all users, and finally the issue that MSN now carries video conferencing which confuses the evaluation. In the CRESS environment, where cost is not a factor for the individual user and resources are available for all collaborators, this level playing field will mean that the chat facility would not be expected to be as highly used as audio and video. It is therefore concluded that if file size is no object the chat facility will not be needed.

The asynchronous communication tools will however provide an additional benefit for those times when an appointment with other collaborators cannot be made. The forum or bulletin board can maintain discussion on particular themes or threads which allows collaborators time to think between posting ideas. Email is a universal tool which, though connected is still outside the CRESS environment and does not need to be incorporated. However, if users felt it more convenient, a link button could be incorporated in the interface to launch the email client.

Message board/News announcements would be particularly useful to supervisors and administrators. This is an element which could be incorporated in the first prototype. Instant messaging is considered to be too distractive an element to be incorporated into the first prototype. However, this needs to be kept under review so as not to limit the interface and rule out a degree of functionality which some users might find useful.

**Identification Tools**
Identification tools are an essential component of communication. There are a number of elements which receive automatic identification when groups meet together face to face but which have to be engineered into the interface when people are meeting online. These include a participant's presence online (logged in), their personal data (name, position etc.), a **focus indicator** (that declares whether they are talking). In addition participants can be represented by avatars depicting images of the participants. A location identifier is sometimes used (particularly in 3D environments).

Identification can be made on three levels. At the lowest level that representation can be a simple name as a presence indicator. At the next level presence may be indicated by a still picture to enable immediate recognition. At the highest level an avatar may be used as a representative within the virtual environment which may include a 3D world. Avatars provide more than a graphical representation and may indicate emotions and other non-verbal communication such as gestures, body language etc. As a 3D or virtual world will not be used in the CRESS environment, avatars will not be considered a priority. However a still picture of the participants will add to the communication and recognition of participants and may be useful for the CRESS environment.

Presence and Focus indicators were perceived not to have disadvantages and these will also be included. Participant's data would also be required to differentiate between, students, supervisors and administrators as well as an indication of their IP address and geographical location. Location identifiers within a 3D environment, would not be required in the CRESS environment.

## Scheduling Space Tools

| Description of Tool | Advantages | Disadvantages |
|---|---|---|
| Scheduling Tools (Calendar) | o Facilitates setting up of online meetings<br>o Allows Collaborators to show availability | |
| Task Setting | o Allows supervisors and others to set timetable of activities and deadlines | |
| Task Monitoring | o Allows all participants to view ongoing progress<br>o Amount of task completion<br>o Can be charted | |

**Table 4 Critical evaluation of Scheduling space tools**

These enable meetings to be set to facilitate the online synchronous communication. It is also used to provide individual task setting and monitoring to enable progress on joint work to be checked and validated. These will not only facilitate appointments for synchronous discussion but also enable tasks to be set and monitored. Each of the three tools above has clear advantages and no observable disadvantages. It is therefore recommended that all three items are adopted in the CRESS environment.

## Shared Working Space Tools

| Description of Tools | Advantages | Disadvantages |
|---|---|---|
| Whiteboard | o General area for working allowing a wide range of use<br>o Brainstorming<br>o Discussion,<br>o Summarising of ideas | o Cannot deal with specific needs such as programming (cannot compile)<br>o Primitive method of drawing |
| Collaborative working window | o Dedicated to particular tasks | o Cannot be used for general tasks |
| 3D environment | o Indicates location of participants and artefacts within a 3D world | o High programming, memory overheads<br>o Not always required for collaboration |

**Table 5 Critical evaluation of shared working space tools**

Working spaces are particular to the tasks which are being performed. These will involve a range of different tools tailored to the different working practices and needs. In some circumstances, a simple whiteboard may suffice while in others a dedicated collaborative working window will be needed. The proposed project will be concerned here with generic workspaces. If required specific tools could be added as modules at a later time. It is not clear at this stage whether a whiteboard would be useful in a CSCR environment. However, the whiteboard is the most popular collaborative working space so it is felt that it should be included and is worth investigating this from a user standpoint before dismissing it as a viable CSCR tool.

Dedicated working spaces which are created to handle a specific task will be interesting only to those for whom the specific task will be important. This kind of dedicated working space is best left as an additional feature to be added as a module at a later time for those who have a specific need for it. It would not be required in a generic CRESS environment. In the same way an output window is too specific. A simulated display is also dedicated to a particular process and is not required. 3D environments would be onerous to program without a large programming team and would not serve any general purpose tool in CRESS but may be employed as a dedicated module for a specific environment.

## Product Space Tools

| Description of Tools | Advantages | Disadvantages |
|---|---|---|
| Output window | o Shows results of calculations or programming or the end product of a process (Graph from equation) | o Takes up space that may not be required widely |
| Simulated display | o Shows in diagrammatic form the operation or working of some part of a process (programming) | o Task specific and has no wider user beyond a particular instance |

**Table 6 Critical evaluation of Product space tools**

This category includes those tools which provide an area for displaying an outcome of the work which is done or under development. This may include room for showing the results of a compiled computer program or it may demonstrate graphically the display of some predetermined outputs given a set of inputs such as a binary display or specifically tailored dashboard instrumentation. These would probably be highly customised and research dependant. In general our CRESS environment would have a limited requirement and could be omitted after the first iteration.

Those tools which are specific to CSCL will be considered in tables 7 to 10.

## Reflection Space Tools

| Description of Tools | Advantages | Disadvantages |
|---|---|---|
| Reflective journal | o Personal and private space for individual contributors to record their reflections on the research process | - |

**Table 7 Critical evaluation of Reflection space tools**

One of the key features to emerge from recent pedagogical theory is the importance of personal reflection in the role of learning. The main tool to be adopted to assist this process is a personal journal or log which allows an individual collaborator to look back upon recent advances in knowledge acquisition or changes to their research through the writing up and recording of their personal journey and exploration of new found knowledge.

## Social Interaction space tools

| Description of Tools | Advantages | Disadvantages |
|---|---|---|
| Community Creation | o Allows the construction of private groups focused upon a particular research subject<br>o Facilitates multiple research groups within the interface | o None known as yet |
| Tags (marking Content) | o Allows rapid searching of varied data according to web 2.0 methods | o Communities may not be large enough to allow full use of social tagging |
| Friend (file sharing) | o Set permissions for who may be allowed download and share files | • Theft of ideas |
| Blog (Public + Private) | o Blogging is an important part of social communication<br>o Allows reflective comments as well as public ones | • Theft of ideas |
| RSS feed to centralize data | o Acts as a central gathering section for information publishing for other parts of the web | • Some important sites may not have RSS feeds |

**Table 8 Critical evaluation of Social interaction space tools**

These are tools which encourage the development of communities within and without the CRESS environment and might involve the creation of tags for marking content, friends for sharing, and communities for the concentration of group effort. All of these will be included in the CRESS environment.

## Assessment/Feedback Space Tools

| Description of Tools | Advantages | Disadvantages |
|---|---|---|
| Assessment | • **Mostly used in CSCL**<br>• **Can mark stages within Postgraduate Degrees** | |
| Feedback | • **Essential for monitoring progress** | |

**Table 9 Critical evaluation of Assessment/Feedback space tools**

The student/supervisor relationship is not an equal one. The flow of information between the two will be of a different character, quantity and quality. The nature of the information flow from student to supervisor maybe exploratory and tentative whereas the information flow in the opposite direction maybe regulative and defining. This latter feedback provides the student with the boundaries within which the student needs to work as well as the encouragement and guidance to move forward in the right direction. An appropriate feedback tool is therefore incorporated into the Collaborative Research Environment for Students and Supervisors (CRESS).

## Supervisor Space Tools

| Description of Tools | Advantages | Disadvantages |
|---|---|---|
| Private area for supervisors | • **Allows unfettered discussion** | |

**Table 10 Critical evaluation of Supervisor space tools**

This is a privileged area for supervisors and deals with their own evaluation of the student's work. It may also afford the opportunity for supervisors to discuss the student's work amongst themselves in a private area to which the students have no access. This provides the opportunity for open and honest debate without worrying the student's response to it. This may take the form of a private chat channel or private forum.

Those tools which are specific to CSCR will be considered in tables 11 to 15.

## Knowledge Space Tools

| Description of Tools | Advantages | Disadvantages |
|---|---|---|
| Database of PowerPoint Slides and Notes etc | o Essential to track contributors<br>o Provides information for security gateway | - |
| Database of research contributions | o Tracking and assigning ownership of work done | - |
| Depository | o File space for the uploading of documents and files<br>o Protected area accessible only by the team | - |
| Academic database | o List of key authors and publications in the field | • - |

**Table 11 Critical evaluation of Knowledge space tools**

This space is designed as a depository for finished work prior to publication as well as for the whole range of documents, papers, and research links etc. which provide the underpinning background knowledge for the research that is taking place. This would involve databases which hold the depository and provided an index and full reference capability such as EndNote. Behind the interface there needs to be a mechanism for storing the information. In particular this will encompass a depository for lodging documents, proposals, papers in progress, research links, PowerPoint slides etc. The advantages clearly outweigh the disadvantages for all four tools; therefore all will be incorporated into the CRESS environment

## Privacy Space Tools

| Description of Tools | Advantages | Disadvantages |
|---|---|---|
| Private Space | o Private area for individual work prior to sharing with collaborators | • - |

**Table 12 Critical evaluation of Privacy space tools**

This is the private area for individual research work prior to sharing with other collaborators. This is concerned therefore with work in progress as it evolves over the research project period. Work from here will eventually uploaded into the feedback space where supervisors can review and comment upon it.

## Public Space Tools

| Description of Tools | Advantages | Disadvantages |
|---|---|---|
| Public information space | o  Public area to publish work in progress surveys for feedback | • |

**Table 13 Critical evaluation of Public space tools**

This is a data and information gathering and disseminating area prior to formal publication. The need for this kind of space may arise from recruitment of the public to surveys or the gathering in of opinions and inviting contributions from a wider area.

## Negotiation Space Tools

| Description of Tools | Advantages | Disadvantages |
|---|---|---|
| Peer Review assistance | o  Facility to email draft copies for validation before publishing<br><br>o  Requires a database of peers who have agreed to provide review feedback | • - |

**Table 14 Critical evaluation of Negotiation space tools**

It is sometimes a long and difficult process to arrive at an agreed course of action during the research cycle amongst collaborators with differing views. Negotiation together with arbitration may be required at times to find the way forward. The use of peer evaluation may well be central to this process. Accordingly a negotiation space is expected to provide tools for lengthy detailed argumentation as well as the introduction of peers or arbitrators external to the immediate research group.

## Publication Space Tools

| Description of Tools | Advantages | Disadvantages |
|---|---|---|
| Publishing assistance | o  Automatic uploading of finished contributions to publication and e-print sites | • - |
| Schemas and Templates | o  Provides Formatting and Styles for particular Journal Publication | • |

**Table 15 Critical evaluation of Publication space tools**

The publication of the final paper could not occur until a number of processes have been completed including document checking for style, format as well as content, argument, coherence etc. This can be assisted with the use of schemas and templates and will also certainly involve a peer review process. Following this assistance with specific journal requirements, style sheets, and final submission to the relevant publication channels will be needed.

## Listing the requirements for CRESS

The foregoing analysis has resulted in the determination of the tool requirements for the CRESS environment which are shown in table 16.

| | CATEGORY SPACES | TOOLS | Required | Not Required | Review |
|---|---|---|---|---|---|
| CSCR / CSCL / CSCW | Administration Space (including security tools) | Login | X | | |
| | | Access/authorisation Tools | X | | |
| | | Recording /Replay Facility | X | | |
| | | Instant Messaging Recording | | | x |
| | | Assistive Agent | | X | |
| | | Help Pane | x | | |
| | | Information Link Map | | x | |
| | | Scenario/Control flow Tools | | | X |
| | Communication tools (including Identification tools) | Text/ Chat | x | | |
| | | Audio/Voice | x | | |
| | | Still Picture | x | | |
| | | Video | x | | |
| | | Instant Messaging | | | x |
| | | Forum | x | | |
| | | Message Board/News | x | | |
| | | Avatar (Representations) | | x | |
| | | Presence Indicator/Information | x | | |
| | | Location Identifier | | x | |
| | | Focus Indication | x | | |
| | | Participant Data | x | | |
| | Scheduling tools | Scheduling Tool | x | | |
| | | Task Setting | x | | |
| | | Task Monitoring | x | | |
| | Shared working space | Whiteboard | x | | |
| | | Collaborative Working Window | x | | |
| | | 3D Environment | | x | |
| | Product Space | Output Window | x | | |
| | | Simulations | | x | |
| | Reflection Space | Reflective Journal/Private | x | | |
| | Social Interaction Space | Community Creation | x | | |
| | | Tags (marking Content) | x | | |
| | | Friend (file sharing) | x | | |
| | | Blog (Public + Private) | x | | |
| | | RSS feed to centralize data | x | | |
| | Assessment / Feedback Space | Assessment | x | | |
| | | Feedback | x | | |
| | Supervisor/Tutor Space | Private area for tutors | x | | |
| | Knowledge Space | Contribution Database | x | | |
| | | Academic database | x | | |
| | | Depository | x | | |
| | | PowerPoint Slides/Notes | x | | |
| | Privacy | Private Space | x | | |
| | Public | Public information space | x | | |
| | Negotiation | Peer Review assistance | x | | |
| | Publication | Schemas/Templates | x | | |
| | | Publishing assistance | x | | |

**Table 16 Summary of tools required for deployment in the CRESS environment**

## Summary

The CSCR Domain has been defined in such a way as to enable analysis and design of many specific and individual interfaces to be constructed for a range of collaborative research interfaces. The analysis of the requirements for a specific interface (CRESS) has been considered in detail.

The methodology of Lindgaard *et al* (2006), was followed except that his first-stage, brainstorming session was replaced with a detailed analysis of pre existing environments to identify user interface elements. This has involved the analysis of 13 e-laboratories and three VLEs to determine a range of tools which have been broken down into a set of 14 logical categories. A specific toolset for CRESS has been arrived at which will initially be incorporated into a storyboard for user analysis.

Future work will involve the building of a CRESS environment which will be based upon full usability analysis. Stage two will involve prototyping, initially in storyboard form, which will be submitted to potential users for initial usability feedback. A prototype will be produced from this and handed over to developers for the construction of the user interface package. This will lead onto usability testing to determine the adequacy of the user interface concepts. Once the basic framework has been established specific plug in modules may be incorporated for specific needs by specific groups. Lindgaard *et al* (2006) original methodology called for three iterations of design, prototype and usability test. However they were not able to maintain this in practice. It is envisaged that at least two or three iterations would be required to provide a stable and usable CSCR environment.